\documentclass[11pt]{article}

\usepackage{geometry}
\usepackage{graphicx}
\usepackage{algorithm, algpseudocode}
\usepackage{amsmath, amssymb, amsfonts, amsthm, float}  
\usepackage{enumerate, color, framed, float, multirow}
\usepackage{comment, longtable, caption, subcaption, appendix}
\usepackage[sort,longnamesfirst]{natbib}
\usepackage{setspace, parskip}
\usepackage{placeins}

\allowdisplaybreaks

\newcommand{\ds}{\displaystyle}

\newcommand{\X}{\mathcal{X}}
\newcommand{\Y}{\mathcal{Y}}
\newcommand{\Z}{\mathcal{Z}}

\newcommand{\real}{{\mathbb R}}

\newcommand\numberthis{\addtocounter{equation}{1}\tag{\theequation}}

\theoremstyle{plain}

\newtheorem{example}{Example}

\theoremstyle{remark}

\begin{document}
\title{Understanding Linchpin Variables in Markov Chain Monte Carlo}

\date{\today}

\author{
Dootika Vats \thanks{Department of Mathematics and Statistics,
Indian Institute of Technology Kanpur, \texttt{dootika@iitk.ac.in}}
\and
Felipe Acosta\thanks{
Natera, 
San Carlos, California,
\texttt{acosta.felipe@gmail.com}}
\and
Mark L. Huber\thanks{
Department of Mathematics and Computer Science,
Claremont McKenna College,
\texttt{mhuber@cmc.edu}}
\and
Galin L. Jones\thanks{
School of Statistics,
University of Minnesota, 
\texttt{galin@umn.edu}}
}
\maketitle

\abstract{An introduction to the use of linchpin variables in Markov
  chain Monte Carlo (MCMC) is provided.  Before the widespread
  adoption of MCMC methods, conditional sampling using linchpin
  variables was essentially the only practical approach for simulating
  from multivariate distributions.  With the advent of MCMC, linchpin
  variables were largely ignored.  However, there has been a
  resurgence of interest in using them in conjunction with MCMC
  methods and there are good reasons for doing so.  A simple
  derivation of the method is provided, its validity, benefits, and
  limitations are discussed, and some examples in the research
  literature are presented.}

\section{Introduction} 
\label{sec:introduction}

Modern statistical models are often sufficiently complicated so as to
require the use of simulation for inference.  Since the seminal work
of \citet{gelf:smit:1990}, Markov chain Monte Carlo (MCMC) has become
the default method for doing so, especially in the context of Bayesian
inference. The Metropolis-Hastings (MH) algorithm \citep{metr:1953,
  hast:1970} is a commonly-used MCMC method due to its flexibility,
ease of implementation, and theoretical validity under weak
conditions.  However, it is often challenging to develop effective MH
algorithms, particularly when the target distribution is
high-dimensional or has substantial correlation between components. A
standard approach is to consider component-wise MCMC methods
\citep{john:etal:2013, jone:etal:2014} such as Gibbs samplers or
conditional MH, also called Metropolis-within-Gibbs, perhaps using
data augmentation \citep{hobe:2011, tann:wong:1987}. However,
component-wise approaches can produce Markov chains that suffer from
slow mixing \citep{beli:1998, jona:2017, matt:1993}.

The limitations of standard MCMC methods in modern applications has
brought about a plethora of new approaches in specific statistical
settings. However, our goal is to highlight
an old and now under-appreciated technique using \textit{linchpin variables}, that can often serve to
simplify the sampling process and provides an organizing device for many of the novel sampling methods.

Before the widespread use of MCMC, the only potentially practical, general tool
for sampling from multivariate joint distributions was the conditional
sampling method \citep{devr:1986, john:1986, horm:wolf:ley:der:2004}. Let $f(x,y)$ be a density function on
$\X \times \Y \subseteq \real^{d_1} \times \real^{d_2}$ and $f_{X|Y}$
be the density function of the conditional distribution of $X$
given $Y$. Let $f_Y$ be the density function of the marginal
distribution of $Y$. If sampling from $f_{X\mid Y}$ is
straightforward, then $Y$ is called a {linchpin variable} \citep{huber:2016} since
\begin{equation}
  \label{eq:lv}
f(x,y) = f_{X \mid Y}(x|y)\, f_Y(y). 
\end{equation}
Thus exact samples can be obtained by first simulating $Y \sim f_{Y}$
followed by $X \sim f_{X \mid Y}$.  This idea is easily extended to
the setting with more than two variables through the usual properties
of joint probability functions.

\begin{example}
\label{ex:rosenbrock}
Consider the Rosenbrock (or banana) density on $\mathbb{R}^{2}$
\[
  f(x,y) \propto \exp\left\{ - \frac{1}{20} \left[ 100(x-y)^{2} +
      (1-y)^{2} \right]\right\}.
\]
This has become a popular and useful toy example for illustrating the performance of MCMC methods in highly correlated settings.  In particular, because the contour plots  resemble the shape of a banana, it can be a challenge to implement an effective MH algorithm. Notice that by inspection of the joint density,  $X|Y=y \sim \text{N}(y^{2}, 10^{-1})$ and integrating $f(x,y)$ with
respect to $x$ yields that $Y \sim \text{N}(1,10)$. Hence $Y$ is a
linchpin variable and it is simple to implement conditional sampling.
\end{example}

Often the linchpin density, $f_{Y}$, is complex enough to prevent
direct sampling from it. When it is difficult to sample from $f_{Y}$
directly, it is natural to turn to MCMC methods for doing so, yielding
a so-called \textit{linchpin variable sampler}. Our goal is to present
advantages of using linchpin variable samplers, highlight some
fundamental theoretical properties, and illustrate examples from the
literature where they have been employed successfully.

An obvious potential benefit to the linchpin variable sampler is that
it naturally reduces the dimension of the MCMC sampling problem since the
target density is the marginal $f_Y(y)$ instead of the joint
$f(x,y)$. Also, the linchpin variable sampler can be particularly
effective when $X$ and $Y$ are heavily correlated (as demonstrated in
a motivating example below); and finally, since information on $X$ is
not required to sample $Y$, all post-processing (like thinning) can
first be done on the linchpin variable, before sampling $X$; see
\cite{owen:2017} for guidance on when thinning a Markov chain
simulation might be useful.

\begin{example}  
Consider sampling from a $p$-variate normal distribution with mean
$\mu$ and covariance $\Sigma$:

\begin{equation}
  \label{eq:mvn_split}
  \left( \begin{array}{c} X_1 \\  X_2 \end{array} \right) \sim
        N_p \left(\left( \begin{array}{c} \mu_1 \\  \mu_2 \end{array}
          \right), \,\, \Sigma = \left( \begin{array}{cc} \Sigma_{11} & \Sigma_{12}
                            \\  \Sigma_{21} & \Sigma_{22} \end{array}
                        \right) \right)\,, 
\end{equation}
where $\mu_1 \in \real^{p-r}$ and $\mu_2 \in \real^r$, $r <
p$. The full conditional distributions are
\begin{align*}
X_1 \mid X_2 = x_2 & \sim N_{p-r} \left(\mu_1 + \Sigma_{12} \Sigma^{-1}_{22} (x_2 - \mu_2), \Sigma_{11} - \Sigma_{12} \Sigma_{22}^{-1} \Sigma_{21} \right)\quad \text{and} \\ 
X_2 \mid X_1 = x_1 & \sim N_{r} \left( \mu_2 + \Sigma_{21}
                     \Sigma_{11}^{-1}(x_1 - \mu_1), \Sigma_{22} -
                     \Sigma_{21} \Sigma_{11}^{-1} \Sigma_{12}\right)
                     \,. 
\end{align*}

Let $p = 5$, $r=1$, and $\Sigma$ be the $5 \times 5$ autocorrelation
matrix with autocorrelation $\rho \in \{.5, .99\}$.  MCMC algorithms
are easily implemented in this example.  For example the above full
conditionals make it easy to implement a Gibbs sampler while a
linchpin variable sampler with the linchpin variable being $X_2$ is
also straightforward. Here, for the marginal of $X_2$, consider an MH
algorithm with proposal $\text{Uniform}(x_2 - h, x_2 + h)$ with $h$
chosen to yield the approximate optimal scaling of
\cite{rob:gel:gilks}.

Starting from the origin, both samplers are run for 5000 steps. The
results are given in Figure~\ref{fig:mvn}. When $\rho = .5$, both
methods perform similarly; however, as expected
\citep[cf.][]{raft:lewi:how:1992}, when $\rho = .99$, the Gibbs
sampler suffers from slow convergence. The linchpin variable sampler
is unaffected by the higher correlation in the target distribution, as
this correlation does not affect the marginal distribution for $X_2$.
\begin{figure}[htbp]
  \centering
  \includegraphics[width=0.4\textwidth]{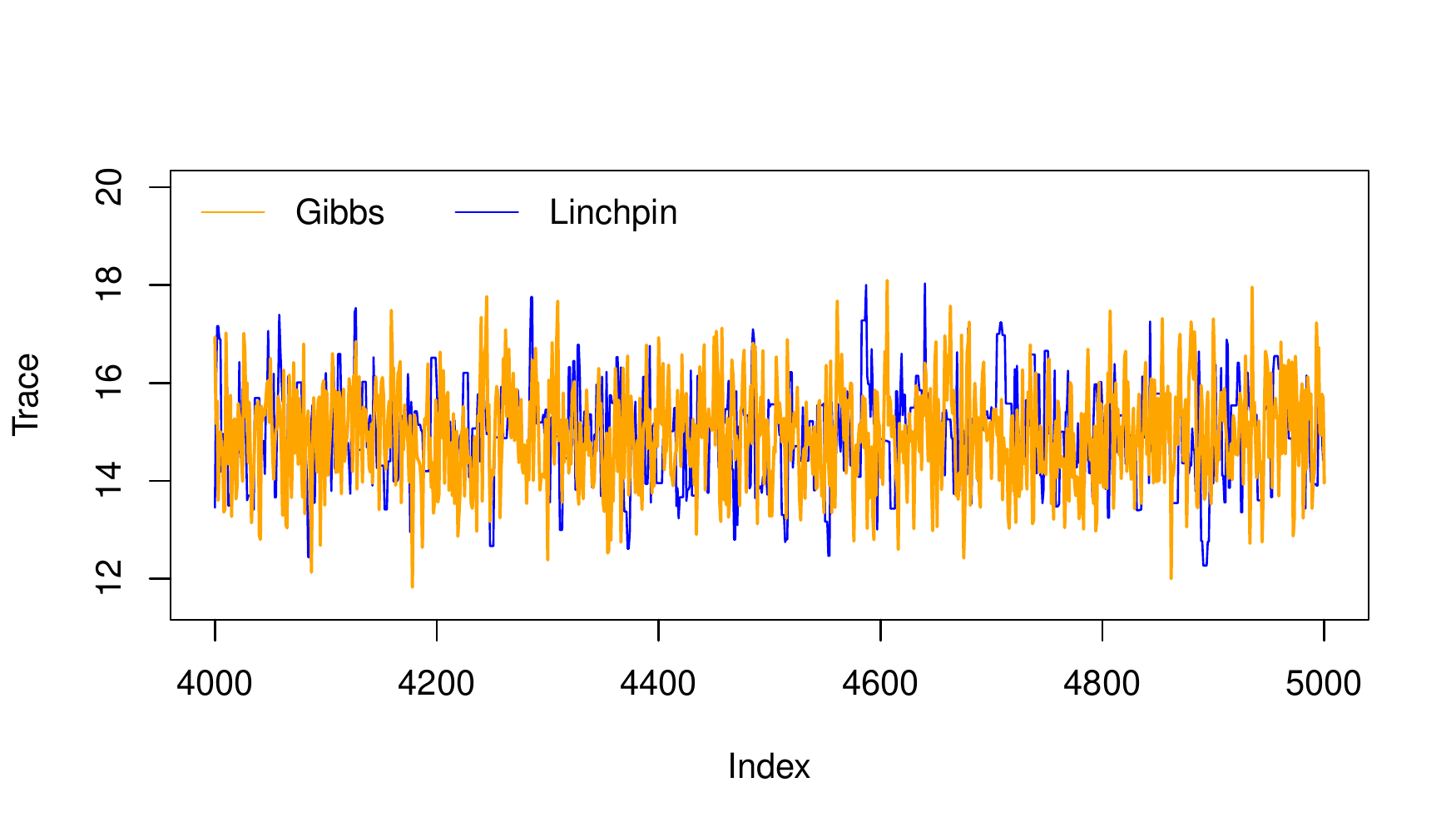}
  \includegraphics[width=0.4\textwidth]{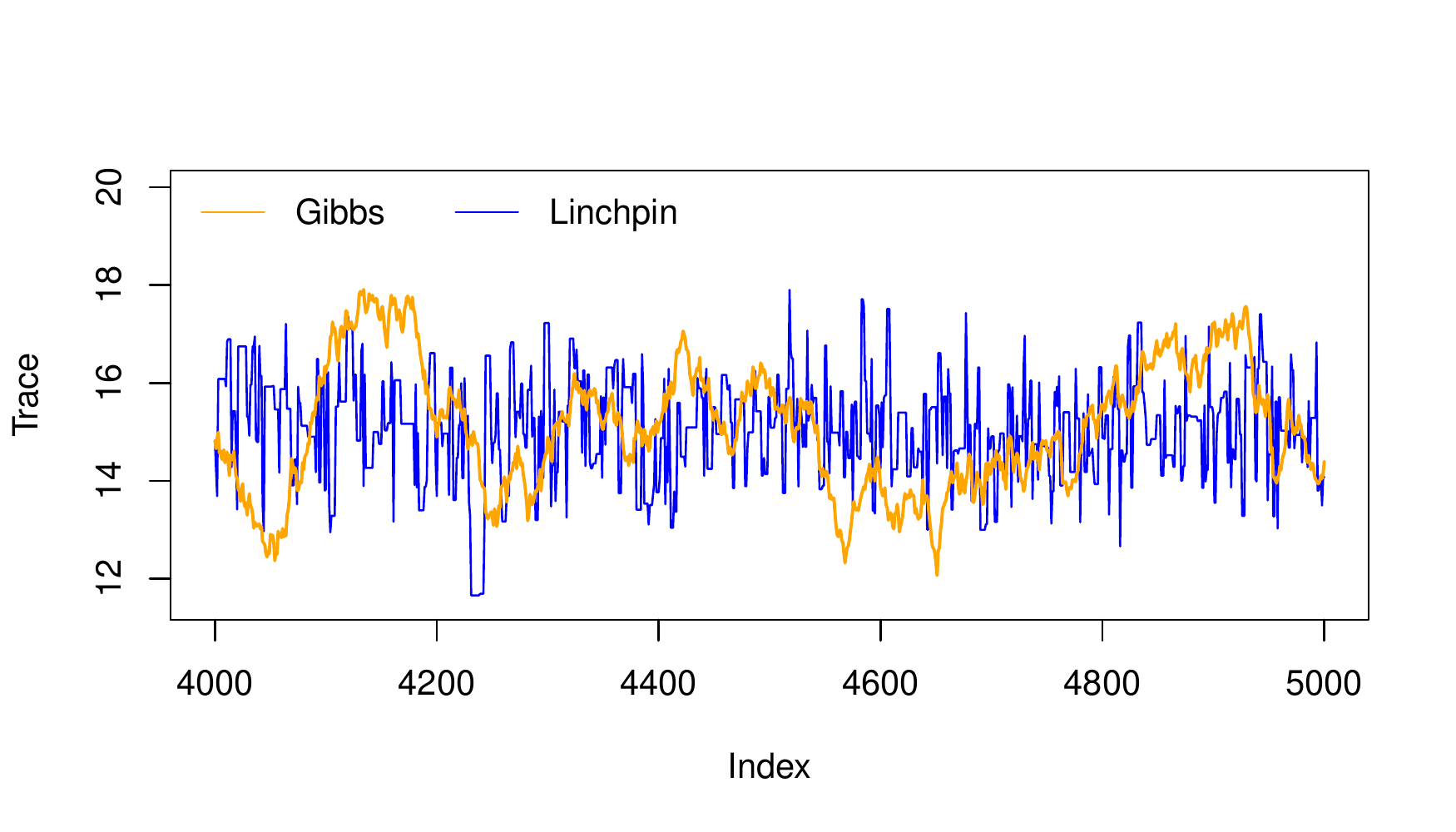}
  \caption{Trace plot for the last 1000 samples for $\rho = .50$ (left) and $\rho = .99$ (right)}
  \label{fig:mvn}
\end{figure}

\end{example}

\section{Linchpin variable sampler}
\label{sec:lv}

Linchpin variable samplers yield valid MCMC algorithms and provide an
organizing principle for seemingly disconnected Monte Carlo methods,
but some basic MCMC concepts are required to get to that point.

\subsection{Fundamentals of MCMC}
\label{sec:stability}

A typical goal of using MCMC methods is to estimate features of a
specified target density.  For example, suppose $F$ is a probability
distribution having support $\mathcal{Z}$ and density $f$.  If
$h : \mathcal{Z} \to \real$, the expectation
\[
\mu := \text{E}_{F}[h(Z)] = \int_{\mathcal{Z}} h(z) f(z) dz\,,
\]
may be of interest\footnote{Often there are several expectations of
  interest, along with quantiles, marginal density plots, and so on,
  but the focus here is on estimating a single expectation; for the
  more general setting see \cite{robe:etal:2021} and
  \citet{vats:fleg:jones:2019}.}.  In many applications, the
complexity of $f$ makes $\mu$ analytically intractable, forcing a turn
to sampling methods for its estimation.

For any MCMC experiment, the practitioner is faced with two
fundamental practical issues: (1) assessing when the sampling
algorithm produces useful observations and (2) using the observations
to reliably estimate $\mu$, that is, another MCMC experiment of the
same run length will produce a similar estimate of $\mu$.

MCMC algorithms simulate realizations of a time-homogenous Markov
chain, $Z_1, Z_2, Z_3, \ldots$, whose dynamics are given by a Markov
transition kernel.  Informally, the reader can think of the kernel as
giving the probability of moving to a set $A$ in $n$ steps given that
the current state is $z$, or
\[
P^n(z, A) = \Pr (Z_{j+n} \in A \mid Z_{j}=z)~~~~~j, n \ge 1.
\]
For simplicity of exposition we will restrict discussion to Markov
chains that have a transition density\footnote{This should not be
  viewed as a real limitation since all of the arguments herein apply
  naturally, with the appropriate adjustments, much more generally.},
that is, suppose $k : \Z \times \Z \to [0, \infty)$ so that
\[
 \int_{\Z} k(z' \mid z) dz' =1 ,
\]
and if $P^1 = P$, then
\[
P(z,A) = \int_{A} k(z' \mid z) dz'.
\]

To ensure that the simulation will eventually produce representative
samples from $F$, $k$ should have invariant density $f$, that is, it
satisfies
\begin{equation}
\label{eq:invariant}
f(z') = \int_{\Z} f(z) k (z' \mid z) \; dz \; . 
\end{equation}
This means that if $Z_n \sim F$, then $Z_{n+1} \sim F$ so the Markov chain is stationary.  Of course, in most MCMC experiments it is difficult to make a draw directly from $F$.  A common way of ensuring that $f$ is invariant for $k$ is by constructing it so that it satisfies the detailed balance condition with respect to $f$, that is, 
\begin{equation}
\label{eq:dbc}
k(z' \mid z) f(z) = k(z  \mid z') f(z') ~~~~~~\text{ for all } z, z' \in \Z.
\end{equation}
It is easy to see by integrating both sides that detailed balance implies $f$ is invariant for $k$. 
Gibbs samplers have transition densities satisfying \eqref{eq:invariant} and, in a formal sense, MH samplers satisfy \eqref{eq:dbc}.

Throughout the Markov kernel (chain) is assumed to be aperiodic,
irreducible, and positive Harris recurrent \citep[see][for definitions
and a thorough treatment of the consequences]{meyn:twee:2009}.  These
assumptions are standard and are typically easily met in MCMC
applications \citep[see, e.g.,][]{ robe:rose:2006, tier:1994}. Let
$\|\cdot\|$ be the total variation norm.  The assumptions imply the
Markov chain is {\em ergodic}
\begin{equation}
\label{eqn:ergodicity}
\|P^n(z, \cdot) - F(\cdot) \| \to 0 ~~~ \text{ as } ~~~ n \to \infty.
\end{equation}
This implies that $Z_{n} \stackrel{d}{\to} F$ as $n \to \infty$, i.e.,
the marginal distribution of $Z_n$ converges weakly to~$F$.  Thus, as
the Monte Carlo sample size increases, an MCMC simulation will produce
a representative (albeit dependent) sample from $F$ no matter the
starting point, but the initial distribution can affect how long it takes
for this to happen.

With a representative sample in hand, estimation of $\mu$ is easy
since the assumptions imply that the sample mean converges to $\mu$.
That is, as $n \to \infty$, with probability~1,
\begin{equation}
\label{eqn:slln}
\bar{h}_{n} := \frac{1}{n} \sum_{i=1}^{n} h(Z_i)  \to \mu.
\end{equation}
Note that if the Markov chain strong law holds for one initial
distribution, it holds for every initial distribution, including point
masses.

The rate of convergence plays a crucial role in the finite-time
reliability of simulation experiments. Let $M$ be a nonnegative
function on $\Z$ and $\gamma$ a nonnegative function on
$\mathbb{Z}_{+}$ such that
\begin{equation}
  \label{eq:tvconv}
  \|P^n(z, \cdot) - F(\cdot)\| \le M(z) \gamma(n).
\end{equation}
If $\gamma(n) = \rho^n$ for some $\rho <1$, then $P$ is {\em
  geometrically ergodic}.  Also, if $P$ is geometrically ergodic with
a bounded $M$, $P$ is {\em uniformly ergodic}.  Constructive methods
for establishing the existence of $M$ and $\gamma$ from
equation~\eqref{eq:tvconv} have been applied in many settings, but are
beyond our scope; see \cite{jone:hobe:2001} for an introduction.

The significance of geometric ergodicity largely lies in the fact that
it is a sufficient condition for the existence of a central limit
theorem (CLT); see \citet{chan:geye:1994}.  More specifically, if
$E_{F} [h^{2+\delta}(Z)] < \infty$ and the Markov chain is
geometrically ergodic, then
\[ 
\sqrt{n} (\bar{h}_{n} - \mu) \stackrel{d}{\to} N(0, \sigma^2_h) .
\]
Now $\sigma^2_h$ is complicated since it accounts for the
autocorrelation in the Markov chain.  Fortunately, there are several
methods to estimate it and then use the resulting confidence regions
to assess the reliability of the estimates; \citep[see
e.g.][]{fleg:hara:jone:2008, fleg:jone:2010,
  jone:hara:caff:neat:2006}.

\subsection{Linchpin variable MCMC sampler} 
\label{sec:linchpin_variable_sampler}

Let $\Z = \X \times \Y$ with $Z = (X,Y)$, so that $f(z) = f(x,y)$
where $x \in \X$ and $y \in \Y$. Assume that it is straightforward to
sample $f_{X \mid Y}$, but MCMC is required to do so from $f_{Y}$.
The linchpin variable MCMC sampler is given in
Algorithm~\ref{alg:lvs}, where $k_Y$ is a Markov chain transition
density that keeps $f_Y$ invariant.

\begin{algorithm}[H]
  \caption{Linchpin variable sampler} \label{alg:lvs}
  \begin{algorithmic}[1]
  \State {\it Input:} Current value $(X_j, Y_j)$
  \State Draw $Y_{j+1} \sim k_{Y}(\,\cdot \, |Y_j)$. 

  \State Draw $X_{j+1} \sim f_{X|Y}(\,\cdot\, | Y_{j+1})$.

  \State Set $j=j+1$
\end{algorithmic}
\end{algorithm}

In Algorithm~\ref{alg:lvs}, $k_Y : \Y \times \Y \to [0,\infty)$ so that it is a Markov transition density with invariant density $f_{Y}$.  That is,
\[
\int_{\Y}k_Y(y'\mid y) dy' = 1 \quad \text{and } \quad f_{Y}(y') = \int_{\Y} f_{Y}(y) k_Y(y'\mid y) dy .
\]
Consequently, the Markov transition density for the linchpin variable sampler
$(x,y) \mapsto (x', y')$ is
\begin{equation}
\label{eq:mtd_linchin}
k(x', y' \mid x, y)  = f_{X\mid Y}(x' \mid y') \,k_Y(y' \mid y) .	
\end{equation}
Notice  that $k$ leaves $f$ invariant since
\begin{equation*}
  \begin{split}
    \int \int k(x', y' \mid x, y) f(x,y) \, dx \, dy &= \int \int
    f_{X\mid Y}(x' \mid y') \,k_Y(y' \mid y) f(x,y) \, dy \, dx\\ 
    & = f_{X\mid Y}(x' \mid y') 
    \int\,k_Y(y' \mid y) f_{Y}(y) \int f_{X\mid Y}(x \mid y)\, dx \, dy\\ 
    & = f_{X\mid Y}(x' \mid y') f_{Y}(y')\int k(y \mid y')  \, dy\\
    & = f(x', y') .
\end{split}
\end{equation*}
Also, $k_Y$ satisfies the detailed balance condition when 
\begin{equation}
\label{eq:marginal reversible}
k_{Y}(y' \mid y) f_{Y}(y) = k_{Y}(y  \mid y') f_{Y}(y') ~~~~~~\text{ for all } y, y'.
\end{equation}
Transition kernel $k_Y$ satisfies detailed balance with respect to $f_Y$ if and only
if the linchpin variable sampler satisfies detailed balance
with respect to $f$.  That is, for all
$x', y', x, y$
\begin{equation}
\label{eq:joint reversible}
k(x',y' \mid  x, y) f(x,y) = k(x, y \mid x', y') f(x', y') .
\end{equation}
To see this, first assume \eqref{eq:joint reversible}.  Let $y,y' \in \Y$ and let $x' \in \X$ be such that $f_{X|Y}(x' \mid y') > 0$.  Then
\begin{align*}
 f_Y(y)k_Y(y'\mid y) = & \int_\X f(x,y) k_Y(y'\mid y)\, dx \\
               = & \int_\X \frac{f(x,y)f_{X|Y}(x'\mid y')k_Y(y'\mid y)}{f_{X|Y}(x'\mid y')}\,dx \\
               = & \int_\X \frac{f(x,y)k(x',y'\mid x,y)}{f_{X|Y}(x'\mid y')}\,dx \\
               = & \int_\X \frac{f(x',y')k(x,y\mid x',y')}{f_{X|Y}(x'\mid y')}\,dx \\
               = & \int_\X \frac{f_{X|Y}(x'\mid y')f_Y(y')f_{X|Y}(x\mid y)k_Y(y\mid y')}{f_{X|Y}(x'\mid y')}\,dx \\
               = & f_Y(y')k_Y(y\mid y')\int_\X f_{X|Y}(x\mid y)\,dx \\
               = & f_Y(y')k_Y(y\mid y')\,,
\end{align*}
and hence \eqref{eq:marginal reversible} holds.  Now assume \eqref{eq:marginal reversible} so that,
\begin{align*}
 f(x,y)k(x',y'\mid x,y) = & f(x,y)f_{X|Y}(x'\mid y')k_Y(y'\mid y) \\
                      = & f_{X|Y}(x\mid y)f_{X|Y}(x'\mid y')f_Y(y)k_Y(y'\mid y) \\
                      = & f_{X|Y}(x\mid y)f_{X|Y}(x'\mid y')f_Y(y')k_Y(y\mid y') \\
                      = & f(x',y')f_{X|Y}(x\mid y)k_Y(y\mid y') \\
                      = & f(x',y')k(x,y\mid x',y')\,,
\end{align*}
and hence \eqref{eq:joint reversible} holds.

If $P_{Y}^{n}(y, \cdot)$ denotes the conditional distribution of the marginal Markov chain after $n$ steps, and
$F_{Y}$ denotes the distribution associated with $f_{Y}$, then under the regularity conditions discussed in Section~\ref{sec:stability},  as $n \to \infty$,
\[
\| P_{Y}^{n} (y, \cdot) - F_{Y}(\cdot) \| \to 0 .   
\]
The linchpin construction encourages the intuition that the dynamics
of $k_{Y}$ transfers to the dynamics of $k$; this is indeed the
case. Let $\mathcal{L}$ denote conditional distribution, then it is
clear from the construction of the linchpin variable sampler that for
$j \ge 1$
\[
 \mathcal{L}(X_j,Y_j \mid X_0,Y_0,Y_j)=\mathcal{L}(X_j,Y_j \mid Y_j)
 \; .
\]
That is, $\{Y_{j}\}$ is de-initializing for $\{(X_{j}, Y_{j})\}$. From Corollary 2 in \cite{robe:rose:2001}, the Markov chains
converge to their respective invariant distributions at the same rate.
That is,
\begin{equation}
  \label{eq:same}
  \|P^n((x,y), \,\cdot\,) - F(\,\cdot\,)\| = \|P_{Y}^n(y, \,\cdot\,) -
  F_Y(\,\cdot\,)\|\; .
\end{equation}
As discussed in Section~\ref{sec:stability}, the convergence rate is
important for ensuring reliable simulation efforts.  The significance
of \eqref{eq:same} is that one only needs to study the convergence of
$P_{Y}$, that is, the Markov chain targeting $f_{Y}$; a concrete
example is presented in Section~\ref{sec:var}.

The point of this section boils down to the fact that, by
construction, $P_Y$ is a valid Markov kernel for $F_Y$ if and only if
the linchpin variable sampler is valid for $F$.  Moreover, the
dynamics of both Markov chains are determined by the dynamics of
$P_Y$.

\section{Linchpin in the literature}

Given the historic relevance of conditional sampling methods, linchpin
variable samplers have been employed in a variety of scenarios. Their
success in the examples below is typically due to either (1) superior
mixing in the lower-dimensional space, (2) de-correlation of
components via the linchpin variables or (3) lower post-processing
costs.

The following presents three examples from the literature where
linchpin variable samplers have been employed successfully. In
addition to the models described here, linchpin variable samplers have
been employed by \citet{beze:hugh:jone:2018}, \cite{blei:ng:jor:2003},
\cite{nort:chris:fox}, and \cite{west:wel:gal:2014}.

\subsection{Collapsed Gibbs sampler in Bayesian vector autoregression} 
\label{sec:var}

For $p > 0$, let $X_1, X_2, \dots$ be a $p$-vector of predictors, and let $\epsilon_1, \epsilon_2, \dots$ be independent and identically distributed according to a $N(0, \Sigma)$, where $\Sigma$ is an $r \times r$ positive-definite matrix. Let $B \in \mathbb{R}^{r \times r}$ and for some $q \geq 1$ and $i = 1, \dots, q$ let $A_i \in \mathbb{R}^{r \times r}$ such that,
\[
Y_t = \ds \sum_{i = 1}^{q} A_i^T Y_{t-1} + B^T X_t + \epsilon_t\,.
\]
The process $\{X_t\}$ is independent of $\{\epsilon_t\}$. Let
$A = [A_1^T, A_2^T, \dots A_q^T]^T \in \mathbb{R}^{qr \times r}$, and
let all the data observed until time $K$ be
$D = \{(X_1, Y_1), (X_2, Y_2), \dots, (X_K,
Y_K)\}$. \cite{ekva:jone:2019} consider prior specifications for
$(A, B, \Sigma)$. Specifically, for fixed hyper-parameters
$m \in \mathbb{R}^{qr^2}$, $C$, a $qr^2\times qr^2$ positive-definite
matrix, and $D$, an $r \times r $ positive-definite matrix, the three
parameters are given the following independent priors:
\begin{align*}
  f(\text{vec}(A)) & \propto \exp\left\{ -\dfrac{1}{2} [\text{vec}(A) - m]^T C[\text{vec}(A) - m] \right\}\,, \\
f(\Sigma) &\propto \,\det(\Sigma)^{-a/2} \, \exp \left\{\text{tr}\left( -\dfrac{1}{2} D \Sigma^{-1} \right) \right\}\, \quad  \text{ and }\,, \\
f(B) &\propto 1\, .
\end{align*}
\cite{ekva:jone:2019} provide conditions which yield that $(A, B, \Sigma) | D$ has a proper distribution. All the full conditionals are available in closed form, and thus a three variable Gibbs sampler is possible. However, \cite{ekva:jone:2019} note that since $A|\Sigma, D$ and $\Sigma|D$ are available to sample from,  a collapsed Gibbs sampler can be constructed which transitions from $(A, B, \Sigma) \mapsto (A', B', \Sigma')$ as
\begin{align*}
k(A',B', \Sigma' \mid A, B, \Sigma) & = f(B' \mid A', \Sigma', D) k_L(A', \Sigma' \mid  A, \Sigma)\\
& :=\,f(B' \mid A', \Sigma', D) f(A' \mid \Sigma', D)\, f(\Sigma' \mid A, D)\,.
\end{align*}
Thus the linchpin variable is $(A, \Sigma)$ and a Gibbs sampler is employed for the marginal sampling. As explained in \cite{ekva:jone:2019}, the relatively simple form of the transition density here makes it easier to analyze the rate of convergence of the Markov chain, relative to the three-variable Gibbs sampler.  

Collapsed Gibbs samplers have been employed in a variety of models; see  \cite{blei:laff:2009}, \cite{chat:pach:2005}, \cite{koop:leon:stra:2009},  \cite{kuo:yang:2006}, and \cite{papa:rob:zan:2018} for some examples.

\subsection{Bayesian linear models}

Let $Y \in \real^n$, $\beta \in \real^p$, 
$u \in \real^k$, $X$ be a known $n \times p$ full column rank design matrix, and $Z$ a known $n \times k$ full column rank matrix. Also 
assume that $\max\{p, k\} <n$. Then a Bayesian
linear model is given by the following hierarchy
\begin{equation}\label{model1}
\begin{array}{rcl}
 Y|\beta, u, \lambda_E, \lambda_R & \sim & \textrm{N}_n\left(X\beta+Z u, \lambda_E^{-1}I_n\right) \\
 g(\beta)                            & \propto & 1 \\
 u|\lambda_E, \lambda_R           & \sim & \textrm{N}_k\left(0,\lambda_R^{-1}I_k\right) \\
 \lambda_E                        & \sim & \textrm{Gamma}\left(e_1, e_2\right) \\
 \lambda_R                        & \sim & \textrm{Gamma}\left(r_1,
   r_2\right) \,.
\end{array}
\end{equation}
Assume that $e_1, e_2, r_1, r_2>0$ are known
hyper-parameters. \citet{sun:2001} show that this hierarchy results in
a proper posterior. Let $\xi = (\beta^T, u^T)^T$,
$\lambda = (\lambda_E, \lambda_R)^T$ and let $y$ denote all of the
data.  Then the posterior density satisfies
\begin{equation}\label{posterior-m22}
  f(\beta, u, \lambda_E, \lambda_R\mid y) = f(\xi, \lambda\mid y) =
  f_{\xi|\lambda}(\xi\mid \lambda, y) \, f_{\lambda}(\lambda\mid y)\,.
\end{equation}

It is easy to sample the conditional $\xi | \lambda, y$ and hence
$\lambda$ is a linchpin variable for $f(\xi,\lambda\mid y)$.
\cite{acos:2015} employs a random walk MH algorithm on the linchpin
variable $\lambda$. Note that $(\xi, \lambda)$ is of dimension
$p + k + 2$, but the linchpin variable $\lambda$ is only
2-dimensional. The dramatic reduction in the state-space from
$(p+k+2)$-dimensions to $2$-dimension yields a much more pliable MCMC
procedure.

\cite{acos:2015} also notes that an accept-reject sampler that draws
independent samples from $f_{\lambda}(\lambda\mid y)$ is
possible. However, the accept-reject sampler is computationally
expensive.  So, although a complete Monte Carlo procedure using
accept-reject sampling is impractical, starting from stationarity by
drawing the initial state of the Markov chain from
$f_{\lambda}(\lambda\mid y)$ is certainly possible in this linchpin
variable sampler.

\subsection{Bayesian variable selection}

Let $y$ be a response vector in $\real^n$, $X$ be an $n \times p$
matrix of predictors, and $\beta \in \real_{p_n}$ be the vector of
coefficients. \cite{nari:he:2014} introduced the following Bayesian
shrinkage and diffusing priors model:
\begin{align*}
y|X, \beta, \sigma^2 &\sim N(X\beta, \sigma^2 I_n)\\
\beta_i| \sigma^2, Z_i & \sim \mathbb{I}(Z_i = 0)\,N(0, \sigma^2 \tau^2_{0,n}) + \mathbb{I}(Z_i = 1) N(0, \sigma^2 \tau^2_{1,n})\\
P(Z_i = 1) & = 1 - P(Z_i = 0)  = q_n \qquad \text{ and } \qquad \sigma^2  \sim IG(\alpha_1, \alpha_2)\,, \numberthis \label{eq:model}
\end{align*}
where $\tau^2_{0,n}$ and $\tau^2_{1,n}$ are positive functions of $n$
and $\alpha_1, \alpha_2 > 0$.  The latent variables $Z_i$ are
indicators of whether the $i$th variable is active or not.  The
resulting posterior is proper and \cite{nari:he:2014} proposed a Gibbs
sampler to sample from it using a three-variable Gibbs sampler since
the full conditional distribution $\beta| Z, \sigma^2$,
$\sigma^2 | \beta, Z$, and $Z | \sigma^2, \beta$ are all available in
closed form.

Alternatively, one could use a linchpin variable sampler \citep[see
e.g.][]{yang:wain:jor:2016,zhou:yang:vats:2021}. The joint posterior
distribution of $(\beta, \sigma^2, Z)$ admits the following
decomposition.
\[
f(\beta, \sigma^2, Z \mid y) = f(\beta, \sigma^2 \mid Z, y)\,f(Z \mid y)\,.  
\]
As it turns out, $\beta, \sigma^2 \mid Z, y$ is available to sample
from and thus, $Z$ is a linchpin variable.

The reduction in dimension here is only from $2p + 1$ to $p + 1$,
which is not too advantageous. However, one can run any choice of MH
algorithm to sample from $f(Z \mid y)$. The state space for $Z$ is
finite, and thus an irreducible Markov chain on the linchpin variable
is automatically uniformly ergodic, yielding a uniformly ergodic
Markov chain for the joint posterior.  In fact, the finite state space
allows for a detailed study of mixing time of the algorithm or the use
of locally informed proposals which can lead to superior mixing
properties on discrete state-spaces \citep{yang:wain:jor:2016,
  liang2021adaptive, zanella2020informed, zhou:yang:vats:2021}.

\section{Discussion}
\label{sec:discussion}

Conditional sampling algorithms were common before the advent of
MCMC. Although they are still used, as highlighted in the prequel,
their discussion is somehow commonly absent in the toolkit of all MCMC
algorithms.

The linchpin variable sampler provides a unifying framework for
several MCMC algorithms based on the well-established idea of
conditional sampling. The linchpin variable sample is most effective
when the dimension of the linchpin variable is significantly smaller
than the dimension of the joint variable or when the joint target
distribution exhibits significant correlation, and the linchpin
variable does not retain that correlation structure.

There are, of course, cases when the linchpin variable sampler presents
little advantage over other samplers. For example, consider the
Bayesian lasso model with a posterior in $p+1$ dimensions; here $p$ is
the number of regression covariates. A linchpin variable sampler is
possible to implement here, but the linchpin variable is
$p$-dimensional, yielding little advantage. In fact, introducing a
$p$-dimensional auxiliary variable in the Bayesian lasso model yields
an efficient Gibbs sampler \citep{park:cas:2008}. 

Given any sampling problem, it is worthwhile to assess whether using a
linchpin variable construction is feasible, before continuing on to
more complex algorithms.

\singlespacing
\bibliographystyle{apalike}
\bibliography{mcref}
\end{document}